\documentclass[a4paper,fleqn,usenatbib]{mnras}

\usepackage{newtxtext,newtxmath}

\usepackage[T1]{fontenc}
\usepackage{ae,aecompl}

\usepackage{epsf}
\usepackage{graphics}
\usepackage{color}

\usepackage{graphicx}	
\usepackage{amsmath}	
\usepackage{amssymb}	

\newcommand{\ax}{$\alpha_{\rm x}$}

\newcommand{\auv}{$\alpha_{\rm UV}$}

\newcommand{\aox}{$\alpha_{\rm ox}$}

\newcommand{\swift}{{\it Swift}}

\newcommand{\xmm}{{\it XMM-Newton}}
\newcommand{\chandra}{{\it Chandra}}
\newcommand{\lledd}{$L/L_{\rm Edd}$}
\newcommand{\nustar}{{\it NuSTAR}}
\newcommand{\plm}{$\pm$}
\newcommand{\nodata}{---}
\newcommand{\rb}[1]{\raisebox{1.5ex}[-1.5ex]{#1}}

\title[Extreme X-ray low-state of RX J2317--4422]{Swift, NuStar and XMM-Newton observations of the NLS1 
galaxy RX\,J2317.8--4422 in an extreme X-ray low flux state}

\author[D. Grupe et al.]{
Dirk Grupe$^{1}$\thanks{E-mail: d.grupe@moreheadstate.edu},
S. Komossa$^{2}$,
Luigi Gallo$^{3}$, 
Norbert Schartel$^{4}$,
Michael Parker$^{4, 5}$,
\newauthor
Maria Santos-Lleo$^{4}$,
Andrew C. Fabian$^{5}$,
Fiona Harrison$^{6}$,
Giovanni Miniutti$^{7}$
\\
$^{1}$Department of Earth and Space Sciences, Morehead State University, Morehead, KY 40514, USA\\
$^{2}$Max-Planck-Institut f\"ur Radioastronomie, Auf dem H\"ugel 69, 53111 Bonn, Germany\\
$^{3}$ Saint Mary's University, Department of Astronomy \& Physics, 923 Robie Street, Halifax, Canada, B3H 3C3 \\
$^{4}$European Space Astronomy Centre (ESAC), 
   P.O. Box, 78,  E-28691 Villanueva de la Ca\~nada, Madrid, Spain \\
$^{5}$ Institute of Astronomy, University of Cambridge, Madingley Road, CB3 0HA Cambridge, UK \\
$^{6}$ Cahill Center for Astrophysics, California Institute of Technology, 1216 East California Boulevard, Pasadena, CA 91125, USA \\
$^{7}$ Departamento de Astrofísica, Centro de Astrobiología (CSIC-INTA), Campus ESA-ESAC, Villanueva de la Ca\~nada, 28692 Madrid, Spain
}

\date{Accepted XXX. Received YYY; in original form ZZZ}

\pubyear{2018}

\begin{document}
\label{firstpage}
\pagerange{\pageref{firstpage}--\pageref{lastpage}}
\maketitle

\begin{abstract}
We report the discovery of RX J2317.8--4422 in an extremely low X-ray flux
state by 
the {\sl{Neil Gehrels Swift}} observatory in 2014 April/May.
In total, 
the low-energy X-ray emission dropped by a factor 100. 
We have carried out multi-wavelength follow-up observations of this Narrow-Line Seyfert 1 galaxy. 
Here we present
observations with \swift, \xmm, and \nustar\ in October and November 2014 and further monitoring 
observations by \swift\ from 2015 to 2018. Compared with the beginning of the \swift\ observations in 2005, 
in the November 2014 \xmm\ and \nustar\ 
observation RX J2317--4422.8 dropped by a factor of about 80 in the 0.3-10 keV band.  While the high-state \swift\ observations can be interpreted by a partial covering absorption model with a moderate absorption column density of $N_H=5.4\times 10^{22}$ cm$^{-2}$ or blurred reflection, due to dominating background at energies above 2 keV the low-state \xmm\ data can not distinguish between different multi-component models
  and were  adequately fit with a single power-law model.
 We discuss various scenarios like a long-term change of the accretion rate or absorption as the cause for the strong variability seen in RX J2317.8--4422.    

\end{abstract}




\section{Introduction}

Extremes of AGN variability provide us with a powerful tool of understanding the physics of 
the central engine of Active Galactic Nuclei (AGN). 
While AGN typically vary on all time scales by factors of a few, 
some of these exhibit dramatic drops in their X-ray fluxes by factors that can exceed 100. 
The cause of these transitions into deep minimum X-ray flux states in AGN can be associated with 
(a) dramatic changes in the accretion rate onto the central supermassive black hole (SMBH), 
(b) changes in absorption, or 
(c) changes in (relativistically blurred) reflection of coronal X-ray photons off the accretion disc, 
for instance in response to changes in the lamppost height or luminosity. 
All three of these types of events have been identified in the past.
A good example of an AGN that has exhibited a dramatic change in its accretion rate  
is the Seyfert 1.9 galaxy IC 3599 \citep{grupe95a, brandt95, komossa99, grupe15}. X-ray reflection has been suggested 
to be the cause of the strong X-ray flux changes in AGN which we have examined in the past based on \xmm\ observations of AGN 
such as 
Mkn 335 \citep{grupe08, grupe12, gallo13, komossa14, gallo15, gallo18, wilkins15}
and 1H0707--495 \citep{fabian12}.  
The third explanation of huge flux changes in AGN is absorption, 
in particular partial-covering absorption in 
X-rays \citep[e.g. ][]{komossa97, guainazzi98, gallagher04, risaliti05, bianchi09, mizumoto14, parker14, yamasaki16, zhang17, turner18}.

Perhaps the best example of an extreme X-ray absorption event is the Narrow Line Seyfert 1 galaxy 
(NLS1) WPVS 007 \citep{grupe95b, grupe13}. While the cause of the dramatic X-ray drop in this NLS1 had been a mystery 
for a decade, it became clear from FUSE observations, that this low-luminosity, low black hole mass AGN shows 
extremely strong broad absorption line troughs in the UV \citep{leighly09}. Observations 
by \swift, HST, and \chandra\ suggest that we most likely see the AGN through and above a patchy, dusty torus \citep{leighly15}. 

We have an ongoing fill-in program with the {\it Neil Gehrels Gamma-Ray Burst Explorer Mission}
\swift\ \citep[\swift\ throughout the paper, ][]{gehrels04}
to check on the X-ray flux of a sample of 
AGN \citep{grupe10}. In addition, AGN that go through extreme flux changes have been 
discovered through the \xmm\ Slew Survey \citep{saxton08}.
These programs have led to the discovery of several AGN that undergo  
transitions into deep minimum X-ray flux states or display dramatic flux increases allowing us to trigger \xmm\ and \nustar\ observations. 

Besides Mkn 335 and 1H0707--495, we also successfully observed Mkn 1048, PG 0844+349, PG 0043+039, 
PG 2112+059, PG 1535+547, and HE 1136--2304 \citep[]{parker14, gallo11, kollatschny15, 
kollatschny16, schartel07, schartel10, ballo08, parker16, komossaetal17, zetzl18} with \xmm\ and \nustar. 
One of the most recent AGN discovered as part of this survey was  
RX J2317.8--4422 ($\alpha_{2000}$ = 23h 17m 50s, $\delta_{2000}$= --44$^{\circ}$ 22$^{'}$ 27$^{''}$, 
z=0.134). RX J2317--4422 was first detected as a bright X-ray AGN during the ROSAT All-Sky Survey 
\citep[RASS, ][]{voges99, schwope00} and was found to be an AGN with a very steep X-ray spectrum 
with \ax=2.5\plm0.8 \citep{grupe98}. 
From its optical spectrum it was identified to be a NLS1 \citep{grupe99} with a black hole mass 
of $M_{\rm BH} = 7.5\times 10^6 M_\odot$ (see Sect. 3.4) accreting near the Eddington limit.  
A follow-up observation with ROSAT in 1997 May did not suggest any dramatic changes in its X-ray 
flux \citep{grupe01}. 

\swift\ started observing RX J2317--4422 in 2005 May as a calibration 
target \citep{grupe10}. \swift\ continued to observe RX J2317--4422 several times as part of various 
fill-in programs. While none of these observations
 suggested anything special about this NLS1, an 
observation in April 2014 revealed that it had dropped its X-ray flux dramatically by a factor of 
more than 30 compared to the previous \swift\ observations 
within less than a year. After the discovery of this very low X-ray flux, we performed 
additional \swift\ observations, confirming the low flux state. This confirmation allowed us 
to trigger a 20 ks initial \xmm\ observation in 2014 October
to obtain a preliminary low state spectrum, which then led to an 
additional 100ks ToO 
observation  with \xmm\ in conjunction with \nustar\ in 2014 November. 
These \xmm\ and \nustar\ observations are the main focus of our paper.

The outline of this paper is as follows: in \S\,2 we describe the
data reduction of the \swift\ , \xmm\ and \nustar\ observations.  In \S\,3 we 
present the results from the analysis of the light curves and 
X-ray spectroscopy, and in 
 \S\,4 we provide a discussion of the nature of the X-ray low-state of RX J2317--4422. 
Throughout the paper spectral indices are denoted as energy spectral indices
with
$F_{\nu} \propto \nu^{-\alpha}$. Luminosities are calculated assuming a $\Lambda$CDM
cosmology with $\Omega_{\rm M}$=0.286, $\Omega_{\Lambda}$=0.714 and a Hubble
constant of $H_0$=70 km s$^{-1}$ Mpc$^{-1}$. This results in a luminosity distance $D$=631 Mpc
using the cosmology calculator  by \citet{wright06}. All uncertainties are 1$\sigma$ unless stated otherwise. The Galactic foreground absorption in the direction of RX J2317--4422 has a column density of  $N_{\rm H} = 1.07\times 10^{20}$ cm$^{-2}$ \citep{kalberla05}.
For all statistical analysis we use the R package, version 3.2.4 \citep[e.g., ][]{crawley09}.

\section{Observations and Data Reduction \label{observe}}
\subsection{Swift}
Table\,\ref{swift_log} lists the 
\swift\ observations of RX J2317--4422 starting with the first observation in May 2005. 
The \swift\ X-ray telescope \citep[XRT;][]{burrows05} 
was operating in photon counting mode \citep{hill04}. Source counts were selected in a
circle with a radius of 23.6$^{''}$ and background counts in a nearby 
circular region with a radius of 235.7$^{''}$. The 
 3$\sigma$ upper limits and the
count rates of the detections were determined by applying the Bayesian method by \citet{kraft91}.
Some of the detections allowed a spectral analysis using Cash statistics \citep{cash79}. 
For all spectra we used the most recent response file {\it swxpc0to12s6\_20130101v014.rmf}.
The X-ray spectra were analyzed using {\it XSPEC} version 12.9.1p \citep{arnaud96}.

In case spectra could not be extracted from the data we converted the count rates or upper limits 
with an energy conversion factor of 3.61$\times 10^{-14}$ W m$^{-2}$ (counts s$^{-1})^{-1}$ which 
was derived from the early observations when RX J2317--4422 was in a high state. 
Note that the spectrum most likely changed. However, due to the low count rate of RX J2317--4422 
no spectral information could be derived from the \swift\ XRT data during the low state since 2014.

The UV-optical telescope \citep[UVOT;][]{roming05} 
data of each segment were coadded in each filter with the UVOT
task {\it uvotimsum}.
Source counts in all 6 UVOT filters
  were selected in a circle with a radius of 7$^{''}$ and background counts in
  a nearby source free region with a radius of 20$^{''}$. The background corrected counts were converted into magnitudes and fluxes by using the calibration as described in  
\citet{poole08} and  \citet{breeveld10}.
  UVOT Vega magnitudes and fluxes were measured with the task {\it  
uvotsource}.
The UVOT data were corrected for Galactic reddening
\citep[$E_{\rm B-V}=0.012$; ][]{sfd98}. The correction factor in each  
filter was
calculated according to equation (2) in \citet{roming09}
who used the standard reddening correction curves by \citet{cardelli89}. 

 Note that the 7$^"$ source extraction radius, which is larger than the standard 5$^"$ radius, was necessary because during several observations the \swift\ star tracker was unable to lock onto the target and \swift\ started to drift causing the image to be smeared out. This drift has two major effects: 1) enhanced background in the source extraction region, and 2) possible host galaxy contamination of the central source. 
However, at the redshift of RX J2317--4422
we do not expect the contribution of widely extended host emission at these radii.
A point-like host contribution cannot be excluded, and may be part of the reason why the amplitude of variability is smaller at optical wavelengths (see below). 
We have tested different source extraction radii (7, 5 and 3 arcsec),
  and the difference in the derived magnitudes is less than the standard deviation in each filter of about
  0.15 mag. The smaller extraction radius was accounted for by setting the {\it uvotsource} parameter {\it apercorr} to {\it curveofgrowth}. 
However, we also note that in our optical spectrum of  RXJ2317.8--4422
 \citep{grupe04} no host absorption features are detected, arguing against the dominance of host emission
  in the optical.

\begin{table*}
	\caption{\swift~ Observation log of RX J2317.8-4422}
	\label{swift_log}
	\begin{tabular}{cccccrrrrrrr} 
		\hline
		ObsID & Segment 
		& T-start$^{1}$ 
		& T-stop$^{1}$ 
		& MJD 
		& $\rm T_{\rm XRT}^{2}$ 
		& $\rm T_{\rm V}^{2}$
		& $\rm T_{\rm B}^{2}$
		& $\rm T_{\rm U}^{2}$
		& $\rm T_{\rm UVW1}^{2}$ 
		& $\rm T_{\rm UVM2}^{2}$ 
		& $\rm T_{\rm UVW2}^{2}$ \\
		\hline
56630 & 003 & 2005 May 26 08:08 & 2005 May 26 10:09 & 53516.380 & 1085 & \nodata & \nodata & \nodata & 2445 & \nodata & \nodata \\ 
35310 & 001 & 2006 Apr 18 00:18 & 2006 Apr 18 23:02 & 53843.444 & 7022 & 677 & 677 & 673 & 1343 & 1746 & 2745 \\
      & 002 & 2006 Apr 20 05:18 & 2006 Apr 20 23:04 & 53845.340 & 3212 & 264 & 264 & 264 &  529 &  717 & 1075 \\
91650 & 001 & 2013 May 03 13:41 & 2013 May 03 13:52 & 56415.574 &  185 &  61 &  61 &  61 &  122 &   38 &  244 \\
      & 003 & 2013 Jul 23 05:21 & 2013 Jul 23 15:09 & 56496.413 & 3000 & 266 & 266 & 266 &  539 &  789 & 1079 \\
91886 & 001 & 2014 Apr 20 05:26 & 2014 Apr 21 08:49 & 56767.802 & 1577 &  11 & 137 & 715 &  777 & \nodata & 239 \\
      & 002 & 2014 May 01 16:32 & 2014 May 01 19:58 & 56778.760 & 1086 &  72 & 104 & 104 &  207 & 200 & 415 \\
35310 & 004 & 2014 Sep 24 00:55 & 2014 Sep 24 01:12 & 56924.044 &  944 &  80 &  80 &  80 &  159 & 238 & 319 \\
      & 005 & 2014 Oct 20 10:12 & 2014 Oct 20 18:28 & 56950.594 &  562 &  82 &  82 &  82 &  204 & 249 & 327 \\
      & 008 & 2014 Oct 29 22:48 & 2014 Oct 29 23:01 & 56959.955 &  687 &  59 &  59 &  59 &  119 & 147 & 238 \\
80853 & 001 & 2014 Nov 16 17:44 & 2014 Nov 17 22:43 & 56978.260 & 1644 & 134 & 134 & 134 &  268 & 372 & 536 \\
35310 & 009 & 2015 Apr 15 07:11 & 2015 Apr 15 12:11 & 57127.403 & 4845 & 125 & 125 & 125 & 3501 & 377 & 499 \\                
      & 010 & 2015 Aug 19 21:12 & 2015 Aug 19 21:34 & 57253.889 & 1271 & 105 & 105 & 105 &  209 & 332 & 420 \\
      & 011 & 2015 Aug 24 21:13 & 2015 Aug 24 22:55 & 57258.917 & 1151 & 136 & 136 & 136 &  275 & 446 & 551 \\
      & 012 & 2016 Apr 05 20:20 & 2016 Apr 05 23:55 & 57483.920 & 4500 & 127 & 127 & 127 & 3178 & 373 & 507 \\    
      & 013 & 2016 Dec 20 03:35 & 2016 Dec 20 08:34 & 57742.250 & 3544 &  40 & 117 & 117 &  153 & 122 & 259 \\
      & 014 & 2016 Dec 24 03:21 & 2016 Dec 24 03:41 & 57746.147 &  455 &  96 &  96 &  96 &  192 & 255 & 385 \\
93134 & 001 & 2017 Apr 19 13:02 & 2017 Apr 19 13:13 & 57862.547 &  649 &  52 &  52 &  52 &  103 & 160 & 207 \\
      & 002 & 2017 Apr 25 06:11 & 2017 Apr 25 06:21 & 57868.260 &  569 &  43 &  43 &  43 &   85 & 149 & 172 \\  
35310 & 015 & 2018 May 27 06:38 & 2018 May 27 10:07 & 58265.351 & 4443 & 254 & 362 & 225 &  725 & 634 & 545 \\           
      & 016 & 2018 Dec 28 05:59 & 2018 Dec 28 09:36 & 58480.326 & 4765 & 391 & 391 & 391 &  783 & 1108& 1036 \\
		\hline
	\end{tabular}

$^{1}${Start and end times are given in UT}

$^{2}${Observing time given in s}

\end{table*}

\subsection{XMM-Newton}

\xmm\ \citep{jansen01}
visited RX J2317--4422 twice. The first observation was performed on 2014 October 29 (MJD 56960)
for a total of 16.9 ks (Table\,\ref{xmm_log}).
This observation was an initial observation to check on the state of the source and to obtain a low state spectrum. After this observation confirmed the low state
we triggered a second, 104 ks, observation starting on 2014 November 16 22:48 UT (MJD 56978).
This observation was performed simultaneously with \nustar\ (see below) and \swift. 

The EPIC pn camera \citep{strueder01} was operating in Full Frame mode with a thin UV-blocking
filter. Only PATTERN 0-4 (single and double events)
were accepted for further analysis. Due to high energy background flares, part of the observations had to be discarded leaving net exposure times of 14900 s and 89081 s, respectively. 
The EPIC MOS 1 and 2 cameras \citep{turner01} were operated in FullWindow mode with the thin filters. The exposure times were 16472s and 102954s for MOS 1 and 
and 16441s and 102988s for MOS2, respectively. For the event selection, only PATTERN 0-12 
(single to quadruple events)
were accepted.
In all three instruments
source counts were selected in a circle with a radius of 32$^{''}$. The background counts were selected from a nearby source-free region with a radius of 64$^{''}$.  The spectral data were binned with 20 counts per bin. In order to increase the statistical significance of the spectral fits,   the data of all three instruments 
were analyzed simultaneously 
with {\it XSPEC} version 12.9.1p \citep{arnaud96}.

\begin{table*}
	\caption{\xmm~ and \nustar\ Observation log of RX J2317.8-4422}
	\label{xmm_log}
	\begin{tabular}{lccccrr} 
		\hline
		Mission 
		& ObsID  
		& T-start$^{1}$ 
		& T-stop$^{1}$ 
		& MJD 
		& $\rm T_{\rm obs}^{2}$ 
		& $\rm T_{\rm eff}^{2}$ \\
		\hline
\xmm	& 0740040101   & 2014-10-29 22:30 & 2014-10-30 02:41 & 56960.024 &  16911 &  14900 \\
	& 0740040301   & 2014-11-16 22:48 & 2014-11-18 03:16 & 56978.585 &  90998 &  89081 \\
\nustar & 800001030002 & 2014-11-16 16:51 & 2014-11-18 10:41 & 56978.587 & 148161 \\	
		\hline
	\end{tabular}

$^{1}${Start and end times are given in UT}

$^{2}${Observing time ($T_{\rm obs}$ and the effective exposure time ($_{\rm eff}$) after correcting for times of high background flares are given in s. These are the times schedules and used for the EPIC pn.}

\end{table*}

\subsection{NuSTAR}

RX~J2317--4422 was observed with \nustar\ \citep{harrison13}
for 150~ks from 2014 November 16, 17:25 UT (Table\,\ref{xmm_log}).
We reduced the \nustar\ data using the \nustar\ Data Analysis Software (NuSTARDAS) version 1.5.1 and CALDB version 20150904. We filtered for background flares, and found clean on-source exposure times of 83.6~ks and 83.4~ks for FPMA and FPMB (the two \nustar\ focal plane modules), respectively. This is roughly half the duration of the observation, due to \nustar 's low-Earth orbit. We extracted source photons from a 30$^{\prime\prime}$ circular region, centered on the source coordinates, and background photons from a 100$^{\prime\prime}$ circular region on the same chip. The background is non-uniform across the detectors in this observation, so we position the background region as close as possible to the source region (we choose not to use an annular region, as it would overlap with the gap between chips). We combined the spectra from FPMA and FPMB using the \emph{addspec} FTOOL. The spectrum is background-dominated over the full band, and RX J2317.8-4422 is only marginally detected at a 2.3$\sigma$ level. The 3$\sigma$ upper limit is 0.0013 counts s$^{-1}$ which corresponds to a 3-79 keV flux of 9$\times 10^{-17}$ W m$^{-2}$. 
 per detector. Due to the low number of counts we did not perform a spectral analysis of these data.

\section{\label{results} Results}
\subsection{X-ray and UV Variability}

\begin{figure}
	\includegraphics[width=\columnwidth, viewport=18 10 522 528]{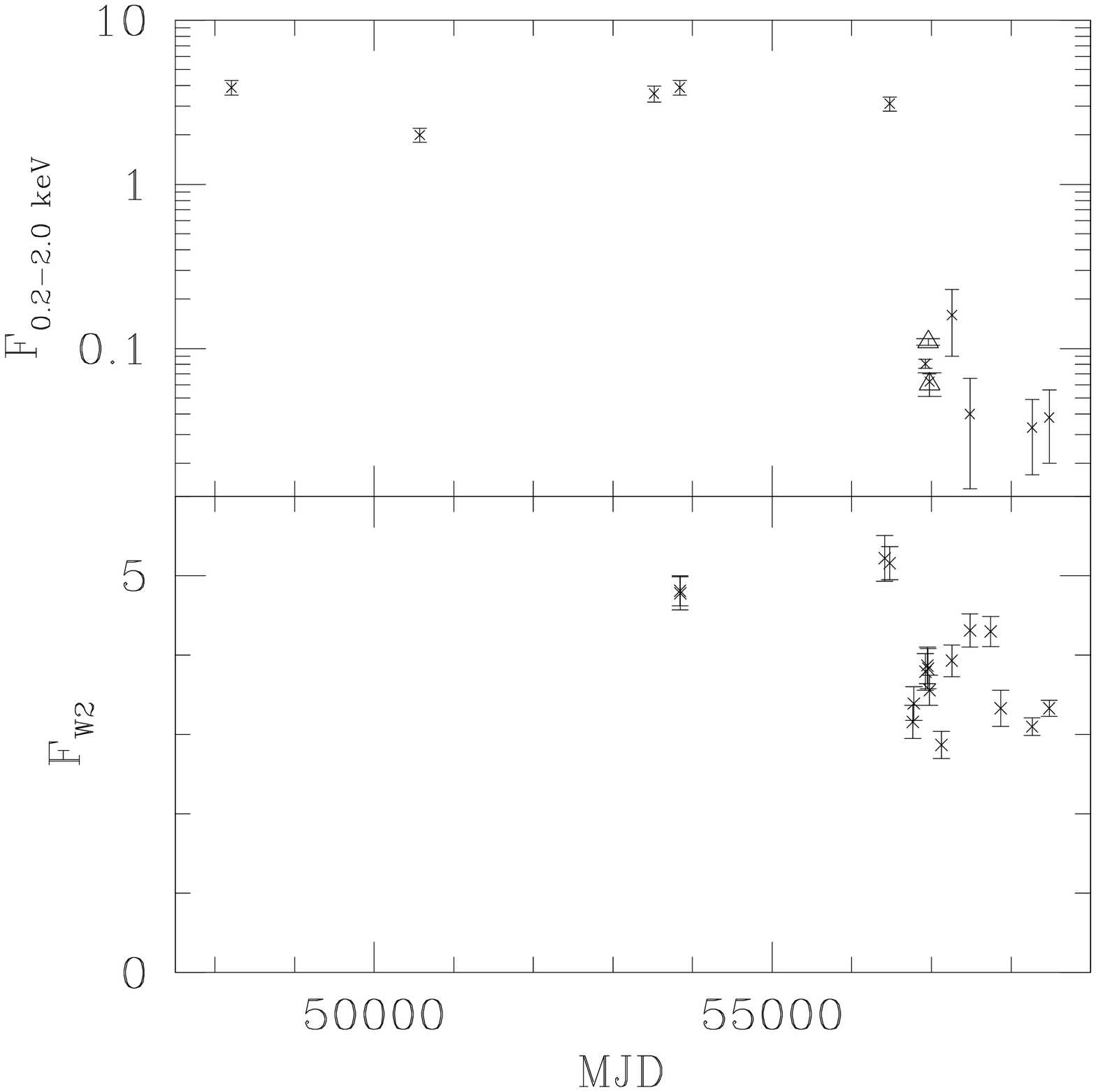}
    \caption{Long-term 0.2-2.0 keV  light curve of RX J2317--4422 corrected for Galactic absorption (upper panel) 
    The first observation was performed during the RASS in November 1990 (MJD = 48208)
    and the second observation by the ROSAT HRI in May 1997  \citep{grupe01} at MJD=50515. 
The triangles  mark the two \xmm\ observations (MJD 56960 and 56978).    
    All other observations were obtained by \swift\ (See Table\,\ref{swift_log}).
The lower panel displays the \swift\ UVOT W2 flux light curve. All fluxes are given in units of $10^{-15}$ W m$^{_2}$.
    }
    \label{rxj2317_xray_lc_long}
\end{figure}

The upper panel of
Figure\,\ref{rxj2317_xray_lc_long} displays the long-term 0.2-2.0 keV light curve of RX J2317--4422 starting with the RASS observation in November 1990 (MJD 48208).
 RX J2317--4422 was only observed once again by ROSAT with the High Resolution Imager in May 1997 (MJD 50575). The observation appeared to be a factor of 2 fainter compared with the RASS observation \citep{grupe01}. However, this is within the normal variability  range of an AGN. RX J2317--422 was not observed again by another X-ray observatory until May 2005 (MJD 53516)
 when it was observed by \swift\ as an X-ray calibration target. All \swift\ observations since May 2005 are listed in Table\,\ref{swift_log}. It was only observed twice again as a fill-in target in April 2006 \citep{grupe10} until we started revisiting RX J2317--4422
  with further guest investigator (GI) 
  fill-in programs to study the long-term X-ray and UV variability of AGN in general. 
Absorption/reddening corrected fluxes in the 0.3-10 keV band (see Sect. 2.1) 
and in all 6 UVOT filters of all \swift\ observations are listed in Table\,\ref{swift_results}. 
Note that the observed X-ray band corresponds to 0.34 -  11.34 keV in the rest-frame. Due to the 
low redshift and relatively low reddening in the direction of the AGN no k-correction was applied. 
While it still appeared to be in its "normal" state during the observation on 2013-July 23, it 
dropped significantly by a factor of at least 8 by 2014 April/May where only a 3$\sigma$ upper limit 
was determined at a level of 2$\times 10^{-16}$ W m$^{-2}$. 
Another \swift\ observation in 2014 September 
suggested a drop by  factor of 12 when compared with the last detection in 2013 July. This drop by a 
factor of $>$10 motivated
us to trigger pre-approved \xmm\ and \nustar\ observations.  The flux levels of the two \xmm\ observations in the 0.2-2.0 keV band are displayed as open triangles in Figure\,\ref{rxj2317_xray_lc_long}.  
The 0.3-10 keV fluxes during the \xmm\ observations were 
about $7\times 10^{-17}$ W m$^{-2}$ and 2.5$\times 10^{-17}$ W m$^{-2}$, respectively. During both observations, the count rate in the \swift\ XRT was too low to obtain a detection.
During most low-state observations
\swift\ was unable to detect the AGN and only 3$\sigma$ upper limits were obtained. 
 Nevertheless, we got \swift\ XRT detections several times including the most recent observation on 
2018-May-27 and December-28
when it was found at a level of about
$3\times 10^{-17}$ W m$^{-2}$ in the 0.3-10 keV band.

\begin{figure}
	\includegraphics[width=\columnwidth, viewport=18 10 522 528]{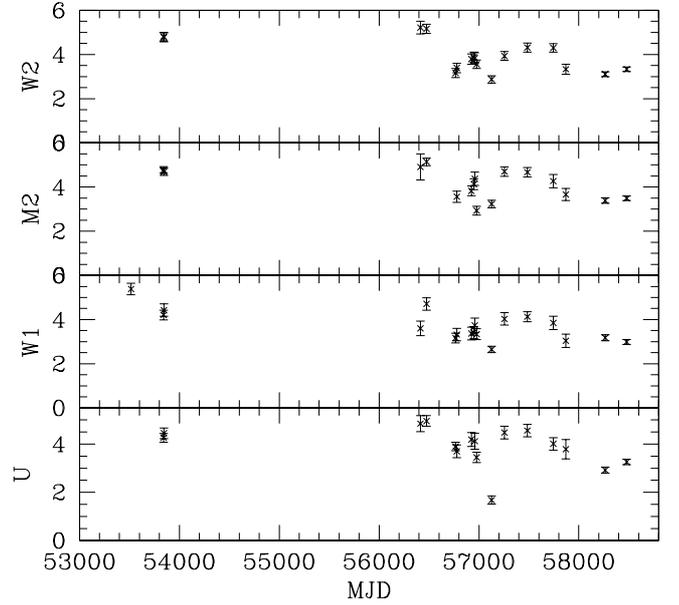}
    \caption{UVOT UV light curves of RX J2317--4422. 
    All fluxes are given in units of $10^{-15}$ W m$^{_2}$ and are listed in Table\,\ref{swift_results}.
    }
    \label{rxj2317_uv_lc}
\end{figure}

\begin{figure}
	\includegraphics[width=\columnwidth, viewport=18 230 522 528]{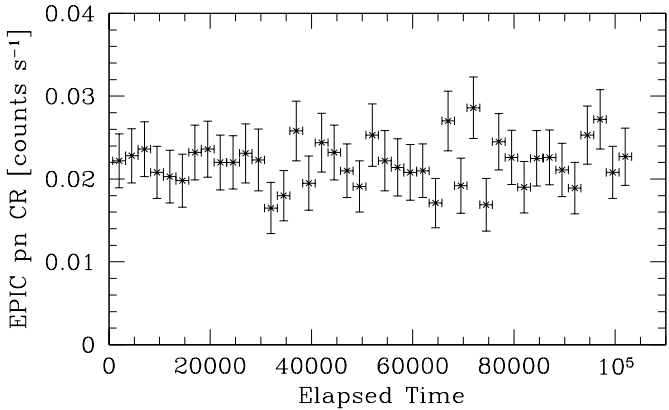}
    \caption{\xmm\ EPIC pn 0.2-2.0 keV light curve during the 2014 November observation using a bin size of 2000s.}
    \label{rxj2317_xmm_pn_lc}
\end{figure}

The UVOT W2 light curve is displayed in the lower panel of Figure\,\ref{rxj2317_xray_lc_long}. The light curves of the  UVOT U, W1, M2, and W2 filters are shown in Figure\,\ref{rxj2317_uv_lc}.
 Table\,\ref{swift_results} also lists the mean, median and standard deviation values of the fluxes in the 0.3-30 keV band as well as all 6 UVOT filters. Clearly the strongest variability can be seen in X-rays, while the variability
  in the UVOT filters is lower, but still significant. The standard deviation in magnitudes is about 0.15 mag in each filter. 
While the long-term as well as the \swift\ XRT and UVOT light curves suggest strong variability in RX J2317--4422, the short-term 
light curve obtained during the 100 ks \xmm\ pn observation in November 2014 shows constant emission.
The \xmm\ pn light curve is shown in Figure\, \ref{rxj2317_xmm_pn_lc} using a bin size of 2000s.

\begin{table*}
	\caption{\swift~ X-ray  and UVOT Fluxes$^{1}$ of RX J2317--4422}
	\label{swift_results}
	\begin{tabular}{cccccccccc} 
		\hline
		ObsID & Segment 
		& MJD 
		& $F_{\rm X}$
		& V 
		& B 
		& U  
		& UVW1 
		& UVM2 
		& UVW2  \\
		\hline
56630 & 003 &  53516.380 & 3.14\plm0.34 & \nodata & \nodata & \nodata & 5.38\plm0.26 & \nodata & \nodata \\ 
35310 & 001 &  53843.444 & 2.11\plm0.09 & 4.22\plm0.19 & 4.00\plm0.13 & 4.26\plm0.18 & 4.21\plm0.22 & 4.77\plm0.15 & 4.81\plm0.19 \\
      & 002 &  53845.340 & 0.44\plm0.13 & 4.43\plm0.28 & 4.25\plm0.19 & 4.45\plm0.22 & 4.43\plm0.28 & 4.69\plm0.17 & 4.78\plm0.21 \\
91650 & 001 &  56415.574 & $<$1.13$^2$ &  4.05\plm0.49 & 4.51\plm0.35 & 4.85\plm0.33 & 3.60\plm0.33 & 4.91\plm0.59 & 5.22\plm0.29 \\
      & 003 &  56496.413 & 1.71\plm0.12 & 4.66\plm0.24 & 4.57\plm0.17 & 4.96\plm0.22 & 4.71\plm0.28 & 5.14\plm0.17 & 5.16\plm0.21 \\
91886 & 001 &  56767.802 & $<$0.21$^2$ &  \nodata & 3.38\plm0.23 & 3.90\plm0.18 & 3.16\plm0.22 & \nodata & 3.16\plm0.21 \\
      & 002 &  56778.760 & $<$0.20$^2$ &  3.24\plm0.38 & 3.65\plm0.23 & 3.70\plm0.26 & 3.32\plm0.28 & 3.56\plm0.25 & 3.39\plm0.21 \\
35310 & 004 &  56924.044 & 0.15\plm0.01 &  3.25\plm0.37 & 3.60\plm0.25 & 4.19\plm0.29 & 3.37\plm0.29 & 3.83\plm0.22 & 3.79\plm0.23 \\
      & 005 &  56950.594 & $<$0.37$^2$  &  4.04\plm0.38 & 3.54\plm0.25 & \nodata & 3.41\plm0.28 & 4.13\plm0.25 & 3.87\plm0.23 \\
      & 008 &  56959.955 & $<$0.51$^2$  & 4.05\plm0.46 & 4.71\plm0.32 & 4.12\plm0.33 & 3.74\plm0.33 & 4.37\plm0.30 & 3.83\plm0.25 \\
80853 & 001 &  56978.260 & $<$0.25$^2$  & 3.83\plm0.31 & 3.76\plm0.21 & 3.45\plm0.21 & 3.34\plm0.25 & 2.93\plm0.20 & 3.56\plm0.19 \\
35310 & 009 &  57127.403 & $<$0.10$^2$ & 3.28\plm0.42 & 3.02\plm0.26 & 1.68\plm0.17 & 2.65\plm0.14 & 3.23\plm0.17 & 2.87\plm0.17 \\	
      & 010 &  57253.889 & 0.11\plm0.07 & 4.38\plm0.36 & 4.09\plm0.23 & 4.59\plm0.26 & 3.85\plm0.53 & 4.46\plm0.22 & 4.12\plm0.21 \\
      & 011 &  57258.917 & 0.12\plm0.08 & 4.33\plm0.32 & 4.00\plm0.21 & 4.37\plm0.26 & 3.90\plm0.28 & 4.50\plm0.20 & 3.79\plm0.19 \\
      & 012 &  57483.920 & 0.040\plm0.026 & 4.51\plm0.42 & 3.92\plm0.25 & 4.56\plm0.26 & 4.13\plm0.22 & 4.67\plm0.21 & 4.31\plm0.21 \\	
      & 013+014 & 57744.146 & $<$0.12$^2$ & 4.13\plm0.76 & 4.30\plm0.30 & 4.01\plm0.22 & 3.85\plm0.22 & 4.27\plm0.30 & 4.30\plm0.19 \\  
93134 & 001+002 & 57865.375 & $<$0.38$^2$ & 3.71 \plm1.01 & 3.42\plm0.44 & 3.79\plm0.40 & 3.04\plm0.30 & 3.66\plm0.27 & 3.33\plm0.23 \\
35310 & 015 & 58265.351 & 0.033\plm0.016 & 3.99\plm0.30 & 3.71\plm0.13 & 2.92\plm0.13 & 3.18\plm0.14 & 3.39\plm0.12 & 3.10\plm0.11 \\
      & 016 & 58480.326 & 0.038\plm0.018 & 2.95\plm0.17 & 2.85\plm0.14 & 3.26\plm0.11 & 2.99\plm0.11 & 3.49\plm0.10 & 3.33\plm0.10 \\
		\hline
\multicolumn{3}{c}{mean} & 0.714 & 3.921 & 3.840 & 3.911 & 3.967 & 4.109 & 3.926 \\
\multicolumn{3}{c}{median} & 0.120 & 4.045 & 3.760 & 4.065 & 3.505 & 4.200 & 3.830 \\
\multicolumn{3}{c}{standard deviation $\sigma$} & 1.088 & 0.509 & 0.528 & 0.813 & 0.688 & 0.677 & 0.731 \\		
		\hline
	\end{tabular}

$^{1}${All fluxes are given  in units of 10$^{-15}$ W m$^{-2}$.      
The UVOT fluxes are corrected for reddening with $E_{\rm B-V}$=0.037 given by
\citet{sfd98}. The X-ray fluxes are in the 0.3-10 keV range and are corrected for Galactic absorption \citet{kalberla05}.}

$^{2}${3$\sigma$ upper limit determined using the Bayesian method as described in \citet{kraft91}.}

\end{table*}

\subsection{X-ray spectral analysis}

\subsubsection{High State 1990-2013}
During the RASS observation in 1990 November the X-ray spectrum in the 0.1-2.4 keV band appeared 
to be very steep with an energy spectral index of \ax=2.5\plm0.8 \citep{grupe01}. This spectral 
slope is consistent with those obtained from the  \swift\ observations in 2005 May and 2013 July 
with
\ax=2.34\plm0.35 and \ax=2.34\plm0.29, respectively.  All these observations indicate a very steep 
X-ray spectrum during the high state consistent with the AGN operating at a high \lledd\ ratio  \citep[see also ][]{grupe10}.  

In order to obtain a higher significance of the fit, we combined all \swift\ data during the high state 
from 2005 to 2013
which resulted in a total exposure time of 14.4 ks. 
Due to the relatively low number of counts (about 180) we did not bin the spectrum and instead
applied Cash statistics \citep{cash79} when fitting the data. All spectral fits were performed with the absorption parameter at z=0 fixed to the Galactic value.  All spectral fit results are summarized in the upper part of Table\,\ref{xray_spec}. 

\begin{figure}
	\includegraphics[width=\columnwidth, viewport=0 30 650 600]{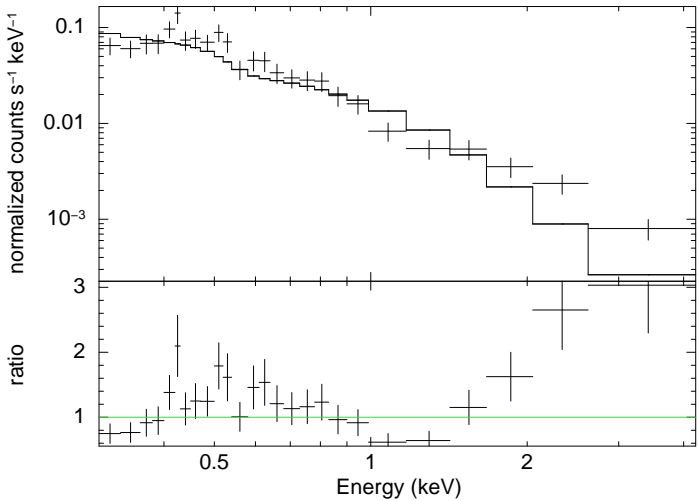}
    \caption{Combined spectrum of all high state data from \swift\ fitted with a single power law model. For display purposes we used the 
spectrum that was binned with 20 counts per bin.}
    \label{rxj2317_swift_high}
\end{figure}

\begin{figure}
	\includegraphics[width=\columnwidth, viewport=0 30 650 600]{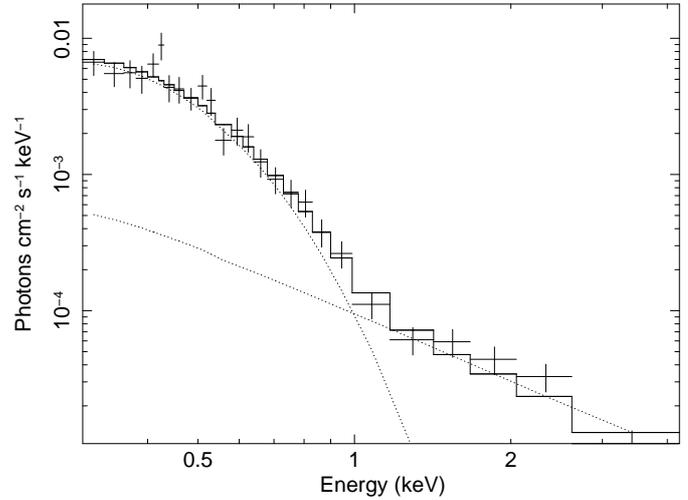}
    \caption{Unfolded combined spectrum of all high state data from \swift\ fitted with a blackbody plus single power law model. Again the binned spectrum shown here was only used for display purposes.}
    \label{rxj2317_swift_high_bb_po}
\end{figure}

We started fitting the data with a simple absorbed power law model which resulted in a steep spectrum with a spectral index \ax = 2.32\plm0.13. However this fit deviates strongly for energies above 1.5 keV as displayed in Figure\,\ref{rxj2317_swift_high} and the C-statistic value over the degrees of freedom is 225/171.
Note that for display purposes we binned the spectrum shown here with 20 counts per bin. Next step was a fit with a combined blackbody plus power law spectrum. This model significantly improved the fit. We found a blackbody temperature of 99 eV and an underlying hard power law spectrum with 
\ax=0.88. The C-stat value here is 169/169. The unfolded spectrum of this fit, again using for display purposes 
a binning of 20 counts per bin, is shown in  Figure\,\ref{rxj2317_swift_high_bb_po}.  Although this is a phenomenological model, it fits the spectrum quite well. 
What this model may suggest is the presence of an underlying starburst.

 A deviation from a simple power law model is typically a sign of the presence of a partial 
covering absorber  or X-ray reflection \citep{gallo06}.
 {\footnote{We checked the background level from 
this combined data set 
and the contribution of background even  at higher energies is negligible. 
The source counts account for 96.6\% of the total number of counts extracted at the source position.}} 
Next, we applied both a redshifted neutral partial covering absorber and an ionized partial 
covering absorber model. Both model fits resulted in similar C-stat values of 194/169 and 198/170, respectively. 
While the neutral 
pc models showed an absorption column density of 5.5$\times 10^{22}$ cm$^{-2}$, 
the absorption column density of an ionized partially-covering absorber was $1.3\times 10^{23}$ cm$^{-2}$. The  ionization parameter $\xi$ was fixed to 10 
(in units of $10^{-9} \frac{W m^{-2}}{m^{-3}}$ or $\frac{ergs s^{-1} cm^{-2}}{cm^{-3}}$). The underlying power law emission spectrum (\ax=2.82) results in a flux in the 0.3-10 keV band of 3.5$\times 10^{-14}$ W m$^{-2}$. 

Often reflection models also 
provide good solutions for spectra like this. Fitting with {\it reflionx}  \citep{ross99, ross05} 
 and the blurred accretion disc model
 {\it kdblur} did  result in a good fit with a C-stat value of 175/173.  
   The partial covering absorber models and the  reflection model produce similar fit results.

\subsubsection{Low State after 2013}

Due to the low X-ray flux state combined with the small effective area of the \swift\ XRT and the relatively short exposure times, none of the \swift\ observations during the low state allowed any kind of spectral analysis. Fluxes were determined by the count rate to flux conversion factor described in section 2.1. 
The spectral information of RX J2317--4422 can be obtained only
from the \xmm\  observations in 2014 October and November.  
For the spectral analysis the absorption column density was set to the Galactic value and the results of all spectral fits are listed in Table\,\ref{xray_spec}. 
The low X-ray flux of RX J2317--4422 in combination with the detector noise in the EPIC pn and MOS cameras made the data above 2 keV significantly background dominated. Therefore we restricted our analysis of the \xmm\ low state data to the 0.2-2.0 keV energy range. As a consequence, complex spectral models like partial covering absorber or reflection models could not be applied.

 For the \xmm\ spectral data analysis we use a combined fit to all pn and MOS 
  spectra where we organized the spectra into two data groups by the October and November 2014 spectra. 
We linked the fit parameters together except the normalizations. 
All results of the spectral fits to the \xmm\ spectra are summarized in Table\,\ref{xray_spec}. Due the the background domination of the data above 2 keV 
no complex spectral analysis appears to be reliable. There is simply no handle on the data at higher energies which define parameters required to describe the partial covering absorber and reflection models. 
The only model, besides the standard single power law model, 
that we applied additionally to the data, was the black body plus power law model which results in  a good fit. This fit is slightly better than that of the single power law model (T-test with T-value = 10.89 and P=0.001). What is interesting to note is that the blackbody component remains basically constant between the October and November observations, while the power law component drops by a factor of about 2.

\begin{table*}
	\caption{Spectral fits High State \swift\ data from 2005-2013 and the 
	 \xmm~EPIC pn and MOS  low state
	spectra. The absorption parameter at z=0 was fixed to the Galactic value, $N_{\rm H} = 1.07\times 10^{20}$ cm$^{-2}$.}
	\label{xray_spec}
	\begin{tabular}{lcccccccc} 
		\hline
		Model 
		& \ax  
		& kT$^1$
		& $N_{\rm H, intr}^2$
		& $f_{\rm pc}^2$ 
		& $\xi^3$ 
		& $Norm_1^4$
		& $Norm_2^4$
		& cstat/dof or $\chi^2/\nu$ \\
		\hline
\multicolumn{9}{c}{High State, \swift, Combined 2005-2013} \\
        \hline
po &  2.32\plm0.14 & --- & --- & --- & --- &   2.42$\times 10^{-4}$ & ---  & 225/171 \\
bb + po & 0.87\plm0.34 & 99\plm7 &  --- & --- &  --- & 2.48$\times 10^{-5}$ & 1.14$\times 10^{-4}$  & 169/169 \\
zpcfabs * po & 2.74\plm0.20 & --- & 5.4$^{+3.0}_{-1.7}$ & 0.87$^{+0.05}_{-0.09}$ & --- & 1.73$\times 10^{-3}$ & ---  & 194/169 \\
zxipcf * po & 2.82$^{+0.17}_{-0.16}$ & --- & 13.7$^{+4.6}_{-4.8}$ & 0.95 & 10 & 4.47$\times 10^{-3}$ & --- & 198/170 \\
po + kdb * refl $^5$ & 1.37$^{+0.16}_{-0.22}$ & --- & --- & --- & 220$^{+377}_{-102}$ & 1.09$\times 10^{-5}$ & 1.44$\times 10^{-7}$ & 175/173 \\
\hline
 \multicolumn{9}{c}{Low State, \xmm\ , 2014-October-30 and November-17} \\
        \hline
   &               &     &     &     &      & 9.69$\times 10^{-6}$ \\
\rb{po} &  \rb{2.65\plm0.16} & \rb{---} & \rb{---} & \rb{---} & \rb{---}  & 5.55$\times 10^{-6}$ & \rb{---}  & \rb{202.7/173} \\
        &               &             &      &      &      & 1.44$\times 10^{-7}$ & 7.94$\times 10^{-6}$ \\        
\rb{bb + po} & \rb{2.61$^{+0.28}_{-0.18}$}  & \rb{122$^{+26}_{-21}$} &  \rb{---} & \rb{---}  &  \rb{---} & 1.27$\times 10^{-7}$ & 4.01$\times 10^{-6}$  & \rb{190/172} \\
		\hline
	\end{tabular}

$^{1}${kT given in units of eV}

$^{2}${The intrinsic absorption column density at the redshift of the AGN is given in units of $10^{22}$ cm$^{-2}$. Note if no uncertainty is given for the covering fraction $f_{\rm pc}$ then it was fixed to the given value. }

$^3$ {$\xi$ is the ionization parameter $\xi = \frac{L}{nr^2}$ given in units of $10^{-9} \frac{W m^{-2}}{m^{-3}}$ or $\frac{ergs s^{-1} cm^{-2}}{cm^{-3}}$. }

$^4${the Normalizations $Norm_1$ and $Norm_2$ for the first and second model are given in units of $10^4$ photons m$^{-2}$ s$^{-1}$ keV$^{-1}$ at 1 keV.
For the low state data, the first row corresponds to the October 2014 \xmm\ data and the second row to the November 2014 spectra.
}

$^5${po + reflionx * kdblur  model. The energy index $\alpha$ for the reflionx model was fixed to the energy index of the power law model.
 For the inner radius we assume a moderately 
rotating black hole with $r_{\rm in} = 2 r_g$.
For the kdblur model the emissivity index was fixed to 5.0 and the inclination angle to 30$^{\circ}$.}

\end{table*}

The 2014 November \xmm\ observation was performed simultaneously with \nustar. 
However, due to the low flux in the NuStar 5-79 keV band we could only obtain a
$3\sigma$ upper limit at a flux level of 9$\times 10^{-17}$ W m$^{-2}$.
 Assuming the power law model found from the \xmm\ data with a spectral slope \ax = 2.65, this corresponds to a flux in the 5-79 keV band of 4$\times 10^{-18}$ W m$^{-2}$, which is consistent with the 3$\sigma$ level derived from the \nustar\ data.

\subsection{UVOT data}
The UVOT W2
light curve is displayed in the lower panel of
Figure\,\ref{rxj2317_xray_lc_long} and it suggests significant variability in the UVOT filters. This variability may suggest color changes between different UVOT bands. We computed B-V, U-B, W2-U, and W2-W1 colors.  The Spearman rank order correlation coefficients of the W2-U and W2-W1 colors are $\rho$=0.63 and 0.56 with probabilities of a random distribution of P=0.015 and 0.028, respectively. 
A correlation analysis of the optical color shows that they are clearly uncorrelated.

\begin{figure}
	\includegraphics[width=\columnwidth, viewport=18 130 550 528]{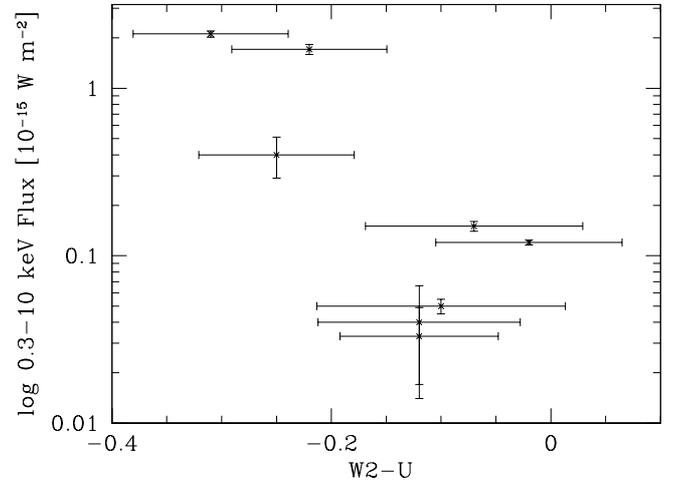}
    \caption{\swift\  UVOT W2-U color vs 0.3-10 keV X-ray flux of RX J2317--4422. }
    \label{rxj2317_xray_w2_u}
\end{figure}

Another question is, are the color changes in  W2-U and W2-W1 colors correlated with the X-ray fluxes. Due to the small number of X-ray detections during the low state, this only allows a very limited analysis. Figure\,\ref{rxj2317_xray_w2_u} displays the W2-U color vs. the 0.3-10 keV X-ray flux of RX J2317--4422. Although this plot suggests that the AGN has a blue UV spectrum when it is brighter in X-rays, due to the small number of data points any correlation analysis will be rather
meaningless. Nevertheless we can apply a 2$\times$2 contingency table. From these data we find that there is a 27\% chance that the AGN is bright in X-rays and blue in the UV and a 73\% chance that is it faint in X-rays and red in the UV. The chances of being faint in X-rays and blue in the UV and bright in X-rays and red in the UV are both 0.

\subsection{Spectral Energy Distribution}

\begin{figure}
	\includegraphics[width=\columnwidth, viewport=18 130 550 528]{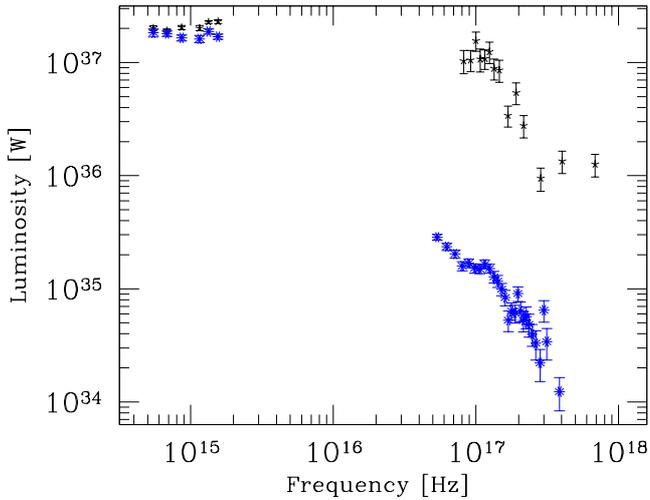}
    \caption{Spectral Energy distributions of RX J2317--4422 during the 2006 high state (black stars) and the 
low state in November 2014 (blue symbols).
}
    \label{rxj2317_sed}
\end{figure}

Figure\,\ref{rxj2317_sed} displays the spectral energy distributions during the high state seen in 2006 
and the low state during the \xmm\ and \nustar\ observation in November 2014. The high state is shown with 
the black stars and and is based on the Swift observations. The low state data from November 2014 are 
using the \swift\ UVOT in the optical/UV and the \xmm\ EPIC pn  data for the X-ray range. 

The high state X-ray spectrum seen in RX J2317--4422 is very soft suggesting a high \lledd\ ratio. 
We went back to the original optical spectrum \citep{grupe04} and re-calculated the black hole mass of 
RX J2317--4422 using the FWHM(H$\beta$) and the luminosity at 5100\AA ~as described in \citet{kaspi00}.
We found that in \citet{grupe04} the black hole mass was underestimated by a factor of two. 
We found that the black hole mass is about 7.5$\times 10^6 M_{\odot}$. 


The optical to X-ray spectral slope \aox\footnote{\aox=$-0.384\times
log(f_{\rm 2keV}/f_{\rm 2500\AA})$; \citet{tananbaum79}} changes from \aox = 1.45 during the 
high state in 2006 to \aox=2.07 in the low state in November 2014. Given the luminosity density at 2500\AA~of 
log($l_{2500\AA}$)=22.24 and following the $l_{2500 \AA}$-\aox relation given in 
\citet{grupe10},
the expected \aox value is 1.36. The \aox\ value during the high state is comparable with the expected value.  
 The UV/Optical part of he SED as shown in Figure\,\ref{rxj2317_sed} appears to be rather flat and may suggest that the optical continuum has a significant host contribution, or is still dominated by the AGN which, however, suffers some extinction.
 The UV spectral slope \auv=2.0 \citep{grupe10} is rather red in particular for a NLS1. This may suggest intrinsic reddening and not a significant contamination by the host. As mentioned earlier, there are no obvious stellar absorption line features in the optical spectrum \citep{grupe04} which also argues from reddening of the AGN emission rather than host galaxy contamination.

RX J2317--4422 is not detected in the Parkes-MIT-NRAO (PMN) radio survey at 4.85 GHz \citep{wright94}. 
Considering that the upper limit in the Southern survey is at 20 mJy (2$\times 10^{-28}$ W m$^{-2}$ Hz$^{-1}$) 
and that the flux in the \swift\ UVOT B-band is of the order of 3.6$\times 10^{-15}$ W m$^{-2}$ or flux density of 0.5 mJy
results in an upper limit of the radio loudness of R$<40$ applying the definition by \citet{kellermann89}.

\section{\label{discuss} Discussion}

\subsection{On the nature of the X-ray low-state of RX J2317--4422} 

While almost all AGN are variable in X-rays, drops by factors $\sim$100 or more are relatively rare.
These strong changes, along with spectral complexity, provide us with important insights in the 
physics of the innermost central region of AGN. Variability and spectral complexity is often strongest in
the class of NLS1 galaxies \citep[review by ][]{gallo18b},
 of which RX J2317--4422 is a member.
The majority of (radio-quiet) AGN with dramatic flux changes has been explained by one of the
following mechanisms:   
\begin{itemize}
\item A change in the accretion rate, for instance through an accretion disc instability or a tidal disruption event.
\item Changes in the effects of (relativistically blurred) reflection of coronal X-ray emission off the accretion disc. 
\item Changes in (neutral or ionized) absorption along our line of sight, partially or fully covering the continuum source.
\end{itemize}
We now comment on (variants of) each of these mechanisms in application to RX J2317--4422. 

\subsubsection{TDE-like accretion event.} 
Most events of stellar tidal disruptions (TDEs) by SMBHs have been observed in non-active galaxies, and
their X-ray emission rapidly fades away on the timescale of months to years \citep[review by ][]{komossa17}. 
Recently, several long-lasting, decade-long events have been identified \citep{lin17}, which
share with RX J2317--4422 the spectral softness at high-state and evidence for near- or super-Eddington accretion. 
However, RX J2317--4422 has been in its bright state for decades, implausibly long
for a TDE, and also shows a classical NLR, indicating long-lasting classical AGN activity. 

\subsubsection{Long-term changes in accretion rate.}
On timescales of years or decades, changes in the accretion rate can lead to fading
of the (X-ray and UV) continuum emission \citep[e.g. ][ and references therein]{noda18, dexter18, lawrence18, ross18}.
In recent years, several such AGN have been found that seem to be slowly switching off, i.e., decreasing 
their accretion rates on the timescale of decades. The best of these cases show a systematic fading
of their broad emission lines, implying a corresponding true intrinsic decrease in the ionizing (EUV) continuum 
emission \citep[e.g. ][]{denney14}.
 RX J2317--4422 likely does not fall in this category, since its 
X-ray light-curve does not show a slow fading, but a rather abrupt drop from a 
(long-lasting) high-state into a (still ongoing) deep low-state.

\subsubsection{Reflection  models.}
When we see a strong hardening of the X-ray spectrum towards higher energy, a possible 
explanation is reflection of X-ray photons on the accretion disc \citep[e.g.,]{ross99, ross05}. 
Although these models can be successfully applied to the \swift\ high-state data, due to the dominating background in the \xmm\ and \nustar\ observations during the low state, no conclusions can be drawn from these data.
A deep long-duration observation with XMM-Newton, as recently carried out for IRAS\,13224--3809 \citep{parker17}
would be required to break model degeneracies, and  constrain the 
reflection (and any other model) components.

\subsubsection{Absorption along our line of sight with or without starburst component.}
Changes in our line of sight (cold or ionized) absorption are known to have strong effects on the 
observed soft X-ray spectra, and can therefore cause high-amplitude X-ray variability.   
High-amplitude absorption variability has been identified in intermediate-type Seyfert galaxies and BAL quasars,
for instance \citep[e.g. ][]{gallagher04, risaliti05, bianchi09}.
In several systems, simultaneous UV and X-ray absorption has been detected 
\citep[e.g. Mkn 1048:][]{parker14, eb16}.
In the extreme case of WPVS\,007, an X-ray drop by a factor of several hundred
was accompanied by extremely strong broad absorption line troughs in the UV \citep{grupe95b, grupe13, leighly15}. 
\citet{gallo06} distinguished between two types of NLS1 X-ray spectra, ``simple systems'' where we have 
a relatively  direct view onto their accretion disc with little
 spectral complexity, 
and ``complex systems''. \citet{jin17} suggested that
those are systems where our line of sight passes
through a clumpy accretion-disc wind. RX J2317--4422 might be in the second class of NLS1 galaxies.
We have tested, whether {\em all} X-ray variability of RX J2317--4422 can be due to absorption. 
For the combined high-state \swift\ data we derived a column density of 
about 5$\times 10^{22}$ cm$^{-2}$. However, due to the low number of source counts above 2 keV during the low state, no parameters of the partial covering absorber could be determined reliably. 

We have also tested, whether the soft X-ray emission during low-state can be due to an additional starburst
component, which is barely detected during high-state, but becomes visible during low-state because of the faintness
of the AGN emission.  
Such a model fits the low-state data well, and NLS1 galaxies are known to show strong
starburst components \citep[e.g.][]{sani10}.
A (0.3-10) keV X-ray luminosity in the starburst 
component of $L= 2.5 \times 10^{34}$ W  
is then required, which is relatively luminous, but still a factor of 10 below the most X-ray luminous 
starbursts \citep[e.g. ][]{komossa03}.


\section*{Acknowledgements}
We thank our referee, Chris Done, for useful comments and suggestions that improved this paper.
We would  like to thank Neil Gehrels and Brad Cenko for approving our ToO requests and
the \swift\ team for performing the ToO observations of
RX J2317--4422. We also want to thank our Morehead State University students Mason Bush, Chelsea Pruett, Sonny Ernst, and Taylor Barber 
for looking through the \swift\ data to find AGN in low X-ray flux states. 
This research has made use of the NASA/IPAC Extragalactic
Database (NED) which is operated by the Jet Propulsion Laboratory,
Caltech, under contract with the National Aeronautics and Space
Administration. 
Based on observations obtained with XMM-Newton, an ESA science mission
with instruments and contributions directly funded by
ESA Member States and NASA.  
This research has made use of the
  XRT Data Analysis Software (XRTDAS) developed under the responsibility
  of the ASI Science Data Center (ASDC), Italy.
ACF acknowledges ERC Advanced Grant 340442.









\bsp	
\label{lastpage}
\end{document}